\documentclass[psfig,epsf]{aa}
\usepackage[dvips]{graphicx}
\usepackage{amssymb,amsmath}

\newcommand\EQ{\begin{equation}}\newcommand\EN{\end{equation}}
\newcommand\EQA{\begin{eqnarray}}\newcommand\ENA{\end{eqnarray}}

\begin{document}

\title{Turbulence in circumstellar disks}
\titlerunning{Turbulence in circumstellar disks}
\authorrunning{Hersant et al.}
\author{F. Hersant$^{1,2,3}$, B. Dubrulle$^{1}$ and J.-M. Hur\'e$^{4,5}$}
\institute{$^1$CNRS URA 2464 GIT/SPEC/DRECAM/DSM, CEA Saclay, F-91191
Gif-sur-Yvette Cedex, France\\
$^2$LESIA CNRS UMR 8109, Observatoire de Paris-Meudon, Place Jules
Janssen, F-92195 Meudon Cedex, France\\
$^3$Institut f\"ur Theoretische Astrophysik, Tiergartenstra\ss e 15,
D-69121 Heidelberg, Germany\\
$^4$LUTh CNRS UMR 8102, Observatoire de Paris-Meudon, Place Jules
Janssen, F-92195 Meudon Cedex, France\\
$^5$Universit\'e Paris 7 Denis Diderot, 2 Place Jussieu, F-75251
Paris Cedex 05, France}
\date{Received ???; accepted ???}

\abstract{
We investigate the analogy between circumstellar disks and the
Taylor-Couette flow. Using the Reynolds similarity principle, the analogy
results in a number of parameter-free predictions about stability of
the disks, and their turbulent transport properties, provided the
disk structure is available. We discuss how the latter can be deduced
from interferometric observations of circumstellar material. We use
the resulting disk structure to compute the molecular transport
coefficients, including the effect of ionization by the
central object. The resulting control parameter indicates that the
disk is well into the turbulent regime. The analogy is also used to
compute the effective accretion rate, as a
function of the disk
characteristic parameters (orbiting velocity, temperature and
density). These values are in very good agreement with experimental,
parameter-free predictions derived from the analogy. The turbulent
viscosity is also computed and found to correspond to an
$\alpha$-parameter $2\times 10^{-4}<\alpha<2\times 10^{-2}$.
Predictions regarding fluctuations are also checked: luminosity
fluctuations
in disks do obey the same universal distribution as energy fluctuations
observed in a laboratory turbulent flow. Radial velocity
dispersion in the outer part of the disk is predicted to be of the
order of $0.1$ km/s,
in agreement with available  observations.
All these issues provide
a proof of the turbulent character of the
circumstellar disks, as well as a parameter-free theoretical
estimate of effective accretion rates.

}

\maketitle

\section{Introduction}
\label{Introduction}

Stars form by gravitational collapse of molecular clouds. During this process,
    proto-stars get surrounded and plausibly fed by the out-coming
    envelope/disk made of gas and dust, which can, under certain conditions,
    coagulate to form planetary embryos. There is little doubt that gas motions,
   usually considered as turbulent, play a major role.
Turbulent motions enhance transport properties, thereby
    accelerate the evolution of the temperature and density in the
    envelope/disk. Also, turbulence may catalyze planet formation thanks to
    the trapping of dust particle inside large-scale vortices (Barge \&
Sommeria 1995; Tanga et al. 1996; Chavanis 2000).
As of now, the assertion that {\it circumstellar disks are
    turbulent} (what we shall refer to as the "turbulent hypothesis'') has
    however never been properly checked. It mainly relies on the fact that
    the luminosity produced by the disk interacting with the central 
star is very
    large (see e.g. Hartmann et al. 1998). In certain cases (FU
Orionis-type systems),
    this luminosity is so high that it even supersedes the stellar
    component.
The most widely accepted scenario so far to account for the abnormal
luminosities of young stellar
objects involves a magnetospheric
accretion for classical T Tauri stars. In this case, the matter in the inner
parts of the disk is coupled with the stellar magnetic field and falls onto
the stars along the field lines at the free fall velocity (see e.g. Gullbring
et al. 1998, Muzerolle et al. 1998). This entails an
accretion shock at the stellar surface in which almost all the visible
and UV excess is produced (veiling the stellar lines). 
This scenario has shown many successes in interpreting spectral features
of T Tauri stars like sodium and hydrogen line profiles (see e.g. Muzerolle et
al. 1998). 
In the case of Fu Ori stars,
the currently
advocated picture involves a wide boundary layer
(see e.g. Popham et al. 1996).
One of the main weaknesses of this scenario is its 
low predictive
power since it all relies on an adjustable parameter (the accretion
rate) which must be postulated {\sl a posteriori} by comparison with
observational data.\\

Indeed, in this framework, the luminosity is directly related
to the amount of
energy released by the disk. 
The
necessity for turbulence comes from the hypothesis
that {\it no laminar motions can produce the amount of energy
dissipated
required to explain observed luminosities} (see e.g. Pringle 1981
and references therein). However, no attempt has ever been made to
substantiate this claim in a quantitative manner. Questions we address here
are: what are the luminosities produced by a laminar disk and by a
turbulent disk ? and how do these compare with observations ?
These two questions are equally difficult to answer, but they hide
different levels of difficulties. One is of theoretical nature: the
physical processes at work in this disk/star interaction region are
complex. A correct description should include simultaneously the
resolution of turbulence (with compressibility effects), radiative
transfer (accounting for UV-irradiation by the star), magnetic
processes, chemistry, the disk flaring, phase separation, time
evolution, etc. The second difficulty is of observational nature. At
the present time,
information has been gathered about the temperature, density and
velocity distributions
{\it in the outer parts of disks}, at $\gtrsim 100$ A.U. typically,
thanks to high resolution interferometry and clever data analysis
(Guilloteau \& Dutrey 1998).
Unfortunately, basic parameters (mean free path, sound velocity and
viscosity) connected with the gas dynamics and dissipation are still
not known {\it in the inner regions}, and especially in the region
where the disk and the star interact.\

Because of these difficulties, we choose to adopt
radically different approach than the classical model: instead of
trying to build a fully
"realistic" circumstellar disk, we use a simplified hydrodynamic
model ("zero order model") and study in detail its physical
properties.
In the future, we will slowly increase its complexity (and reality!) by adding
new ingredients like magnetic field, stratification, radiative
transfer. Here, we show that our zero order model is analog to an
incompressible rotating shear flow. It is therefore amenable to a
simple {\sl
laboratory prototype}, the Taylor-Couette flow. From theoretical and
experimental studies of the properties of this prototype, one can then
build general laws in circumstellar disks by a simple use of the
Reynolds  similarity principle. Taylor-Couette flow is a classical
laboratory flow, and it has been the subject of many experiments. A
recent review about stability properties and transport properties for
use in astrophysical flows has been made by Dubrulle et al (2004b).
As a result, they derive critical conditions for stability, and
simple scaling laws
for transport properties, including the influence of stratification,
magnetic field, boundary conditions and aspect ratio. In the present
paper, we apply these results to circumstellar disks and derive the
expression of the characteristic
parameters of the model as a function of astrophysical observables. We
propose a procedure of quantitative estimate of the observable using
observational results of Guilloteau \& Dutrey (1998) and derive
parameter-free predictions about turbulence and turbulent transport
in circumstellar disks. These predictions are tested against
observational data from T Tauri and FU Ori stars.

\section{Hydrodynamic model}
\label{sec:Hydrodynamic model}

\subsection{Basic ideas}
\label{subsec:model}

\begin{figure}[hhh]
\centering
\includegraphics[width=8.5cm]{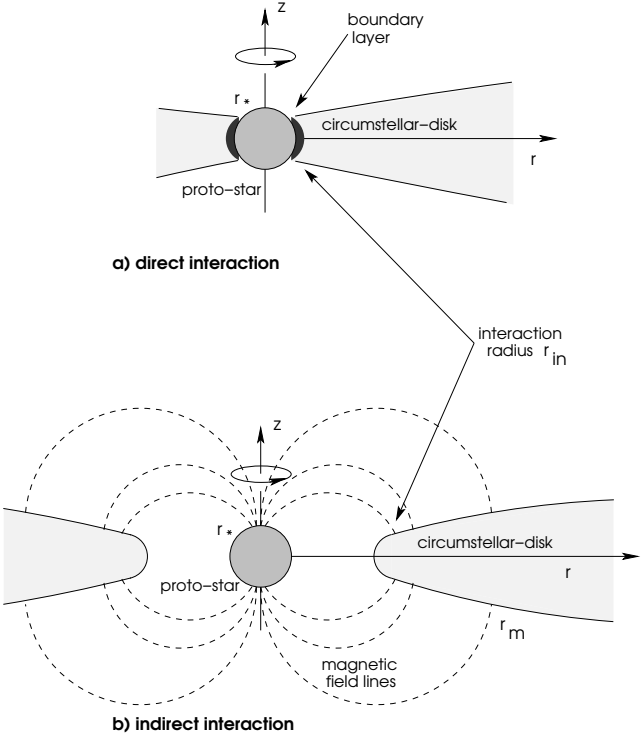}
\caption[]{Two possible configurations considered in the present
model: (a) the whole disk is incompressible and extends onto the
proto-star, and (b) the disk is incompressible until an "interaction
radius" imposed for instance by a magnetic field.}
\label{fig:interaction.xfig.eps}
\end{figure}

Observation of circumstellar disks suggests that they have sizes
between 100 and 1000 astronomical units. In the sequel, we will focus
only on the part of the disk expected to behave like an incompressible fluid.
An estimate of the importance of compressibility can be obtained via
the Mach number, the ratio of the typical velocity to thermal
velocities. It is generally admitted that compressibility effects
start playing a role when this number reaches values of unity. In the
outer part of the disk, this ratio has been estimated by Guilloteau
and Dutrey (1998) from CO line profiles. Its value is about
0.2-0.3. In the inner part of the disk (radius ranging from 1 to 30
astronomical units), we may use the disk structure inferred from
the D/H ratio measured in the Solar System
(Drouart et al, 1999, Hersant et al. 2001), which leads
to a Mach number of the order of $0.05$ to $0.1$. These figures
indicate that both the inner and the outer part can be treated 
as
incompressible fluids. Closer to the star, the situation is less
clear. On one hand, temperature tends to increase strongly, leading
to an increase of the sound velocity and a decrease of the Mach
number. On the other hand, as one gets closer to the boundary, one
may expect larger typical velocities induced by larger velocity
gradients, and thus increase of the Mach number. There are no direct
observations supporting one scenario or the other. We shall then
consider two scenarii: one, in which the Mach number $Ma$ never exceeds
unity. In this case, the whole disk is incompressible, and connects
smoothly onto the star at the star radius. The inner boundary is thus
defined as $r_i=r_\ast$. In a second scenario, the Mach number
reaches unity at some "interaction radius" $r_{\rm in}$, leading to
an inner boundary at $r_i=r_{\rm in}$. At this location, a shock
appears, in which all velocities are suddenly decreased to very small
values. In the shock, all the kinetic energy is transfered to the
thermal energy, thereby producing a strong temperature increase (by a
factor of $1+Ma^2 \simeq 2$). This
entails an increase of ionization and the matter gets more coupled to
the stellar magnetic field.
This second situation is considered in magnetospheric
accretion models (Hartmann et al, 1998, Hartmann et al
2002), in which case $r_{\rm in}$
is the Alfven radius; see Schatzman (1962,1989). Figure
\ref{fig:interaction.xfig.eps} summarizes these two possible
configurations. From a hydrodynamical point of view, in the first
situation the boundary is similar to free-slip boundary (with
possible non-zero velocities in the direction tangential to the star
boundary), while in the second situation, the interaction radius acts
as a no-slip boundary (with all velocities becoming zero). This
difference may reflect in the transport properties, see Dubrulle et
al. (2004b). In the sequel, the free-slip boundary condition will be
referred to as \emph{smooth}, while the no-slip boundary will be
referred to as \emph{rough}.
In the laboratory experiments reviewed in Dubrulle et al (2004b), the turbulent
transport depends on the boundary conditions.
Specifically, transport is enhanced (with respect to any other
boundary conditions) with boundary
conditions of the rough or no-slip type.
In the astrophysical case, it is not quite clear whether these two
boundary conditions apply, or even whether different inner and outer
boundary conditions result into an intermediate transport enhancement.
We shall therefore devise observational tests using quantities
independent of boundary conditions via a suitable
non-dimensionalisation.

\subsection{Basic equations}

In any case,
    the angular velocity $\Omega$ at the inner boundary is that of the star,
    namely $\Omega(r_{i})=\Omega_\ast$. For $r> r_{i}$, the Mach number
of the flow is less than one by construction, i.e. pressure
fluctuations vary over a time scale short compared with the dynamical
time. In such a case, one can assume hydrostatic equilibrium in the
vertical direction, implying a decoupling of the vertical and 
horizontal structure. It is then convenient to describe the disk by 
its
"horizontal equations", obtained by averaging the original equations of
motion in the vertical direction. The procedure is described e.g. in
Dubrulle (1992). It leads to:
\EQA
\partial_t \Sigma+\partial^h_i \Sigma u_i&=&0,\nonumber\\
\partial_t(\Sigma u_i)+\partial^h_j (\Sigma u_i u_j)&=& -\partial^h_i
Hp+\partial^h_j \tau_{ij}-\Sigma \frac{GM}{r^2}e_r,\nonumber\\
\tau_{ij}&=&\mu(\partial^h_i u_j+\partial^h_j u_i)+(\zeta+\frac{\mu}{3})
\partial^h_j u_j.
\label{dubrulle92}
\ENA
Here, $\partial^h$ is the horizontal gradient ($\partial_z=0$), $u$ and $p$ are the
Favre average of the velocity and the pressure over the vertical
direction, $\Sigma$ is the surface density, $\mu$ and $\zeta$ are
surface viscosity coefficients, $G$ is the gravitational constant,
$M$ the mass of the star, and $r$ the distance to the star in
cylindrical coordinate, $e_r$ is a unit vector in the radial
direction and $H$ the vertical scale height. In the hydrostatic
approximation,

\begin{equation}
H=c_s \left( \frac{r^3}{GM}\right)^{1/2},
\label{Hdefin}
\end{equation}

where $c_s$ is the sound velocity.
This expression is only valid when self-gravity can be neglected, namely when:
\begin{equation}
\frac{M_{disk}}{M} \lesssim \frac{H}{R}
\end{equation}
where $M_{disk}$ and $M$ are the masses of the disk and the
star, respectively (see e.g. Hur\'e, 2000).\\
In the opposite case,
$H$ will rather vary like the Jeans length in the vertical direction,
as:

\begin{equation}
H=\frac{c_s^2}{4\pi G\Sigma}.
\label{Hdefinsg}
\end{equation}

These equations should be supplemented with an equation for the
surface energy $E\sim H c_s^2$, but we shall not need it in the
sequel.

\subsection{Stationary axi-symmetric state}

The equation (\ref{dubrulle92}) admits simple basic state, under the
shape of stationary axi-symmetric solution. The mass conservation
then implies:
\begin{equation}
\frac{1}{r}\partial_r (r \Sigma u_r)=0,
\label{mass}
\end{equation}
or
\begin{equation}
u_r=\frac{M_t}{\Sigma r},
\label{circuladisk}
\end{equation}
where $M_t$ is a constant, dimensionally equivalent to an accretion rate.
Plugging this into the radial and azimuthal component of
(\ref{dubrulle92}), we obtain two equations:
\EQA
-\frac{M_t^2}{\Sigma 
r^3}\left(1+\frac{r\partial_r\Sigma}{\Sigma}\right)&-&\frac{\Sigma 
u_\theta^2}{r}=-\partial_r
Hp-\Sigma \frac{GM}{r^2} \nonumber\\
+\frac{2 \mu M_t}{\Sigma r^3}\bigg(-\frac{r \partial_r \mu}{\mu}
\left(1+\frac{r\partial_r\Sigma}{\Sigma}\right)&-&\frac{r^2 \partial_r^2\Sigma}{\Sigma} 
+2 
\frac{(r \partial_r
\Sigma)^2}{\Sigma^2}-\frac{r\partial_r\Sigma}{\Sigma}\bigg), \nonumber \\
\frac{M_t}{r^2}\partial_r( r u_\theta)&=&\frac{1}{r^2}\partial_r\left(\mu
r^3\partial_r \frac{u_\theta}{r}\right).
\label{equathetadisk}
\ENA
The general solution of the second equation of (\ref{equathetadisk}) is :
\begin{equation}
u_\theta=A \exp\left(\int_{r_i}^r (\beta+1) dx/x\right)+\frac{B}{r},
\label{solutheta}
\end{equation}
where $A$ and $B$ are constants and $\beta=M_t/\mu$. Plugging this
solution into the first equation of (\ref{equathetadisk}) then
defines the general pressure.\

\begin{figure}[hhh]
\centering
\includegraphics[angle=270,width=9cm]{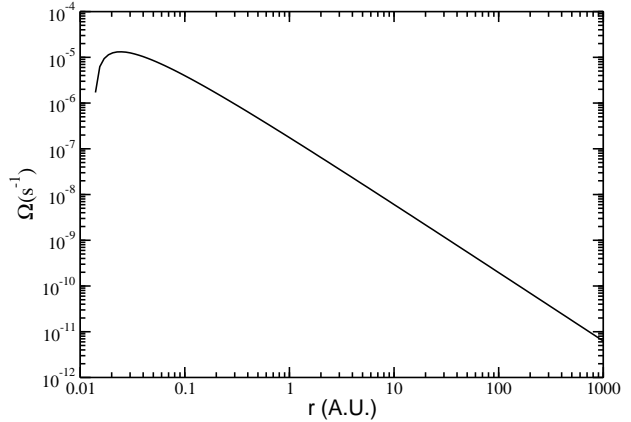}
\caption[]{Velocity profile in a circumstellar disk in the viscous
regime, with keplerian velocity in the outer part, and smooth
matching onto the central object at the interaction radius. For this
example, the mass of the central star has been taken as a solar mass.
}
\label{fig:kepdisk.eps}
\end{figure}

The basic state depends on three constants $A$, $B$ and $M_t$, which
must be specified through some sort of boundary conditions. In the
case of astrophysical flows, the boundary conditions are not very
well known and it is less easy to constrain the parameters. The
condition that the rotation velocity of the disk matches the star
velocity at the interaction radius only provides one relation between
the three parameters:
\begin{equation}
A+\frac{B}{r_i}=r_i\Omega_\ast.
\label{condition1}
\end{equation}
An additional constraint comes from the hypothesis that the disk is {\sl geometrically thin}.
This is consistent with our assumption that the Mach number is less
than unity. In that case, as soon as there is no dramatic variation of the thermodynamic variables, radial pressure gradients and terms involving the radial velocity\footnote{
The viscosity  and the advection terms are then negligible in front of the radial pressure gradient by a factor of the order of $Ma \frac{H}{r}$ and $Ma^2$, respectively.}
 can be neglected in front
of gravity, and the only way to satisfy the first equation of
(\ref{equathetadisk}) is to set $M_t/\mu=\beta=-3/2$.
In this case, the disk is almost Keplerian and obeys:
\EQA
u_r&=&-\frac{3\mu}{2\Sigma r},\nonumber\\
u_\theta&=&\frac{\sqrt{GM}}{r^{1/2}}+\frac{r_i^2}{r}\left(\Omega_\ast-
\Omega_K(r_i)\right),
\label{solkep}
\ENA
where $\Omega_K(r_i)$ is the angular Keplerian velocity at $r=r_i$.
The corresponding solution is plotted in Fig. \ref{fig:kepdisk.eps}.
It is made of a Keplerian disk in the outer part, with a continuous
matching towards the star velocity at the interaction radius.

\subsection{Comment about the disk outer radius}
In the expression we derived for the velocity, we did not need to
specify anything about the condition at the outer boundary of the
disk. One may then wonder what determines this boundary, and whether
it is relevant in specifying the geometry of the problem.
One way to
answer this question is to note that stationary solutions of the
shape ({\ref{solkep}) are only possible provided the dynamical time
scales are short with respect to the viscous time scale. In the
vertical direction, the dynamical time scale to ensure hydrostatic
equilibrium is $H/c_s\sim \Omega_K^{-1}$. In the horizontal plane,
the two dynamical time scale are the radial time scale $r/u_r\sim
r^2\Sigma/\mu$ and the orbital time scale $\Omega^{-1}\sim
\Omega_K^{-1}$. The radial time scale is comparable with the viscous
time scale. So the condition for stationarity is that the orbital
time scale is less than the viscous time scale, resulting in $r<r_o$,
with $r_o$ solution of the equation:
\begin{equation}
r_o=\left (\frac{\mu}{\Sigma \Omega}\right)^{1/2}\vert_{r=r_o}.
\label{defiro}
\end{equation}
This radius defines the outer geometrical limit within which
stationary solutions can reasonably exist. If in this formula, one 
considers the ordinary viscosity, then one typically obtains $r_o$ 
much larger than the observed disk outer radii. On the other hand, 
one may argue that as soon as the disk is turbulent, the molecular 
viscosity becomes irrelevant, one must consider a kind of 
"turbulent viscosity" in this formula. In this case, using the 
formula we derive in Section 3.3, we find that $r_o$ is of the order 
of the disk scale height. In geometrically thin disks, it is not quite clear whether
this limit really exists or not. 
Neither limits really match the observed disk radii. However, the
sharp observed edge of disks remains inconsistent with stationary models.
Stationary solutions, due to constant accretion rate in radius, are
indeed in
essence radially infinite. This suggests that the disk outer radii
may still be linked with some unstationary effects. This is the
subject of a forthcoming paper (Mayer et al. 2004, in preparation).

\subsection{The incompressible analog}

Astrophysical disks are (weakly) compressible and radially stratified.
It is however possible to build an incompressible analog of them,
using clever boundary conditions. This remark is at the heart of the
laboratory prototype. Consider indeed an incompressible, unstratified
fluid, enclosed within a domain with cylindrical symmetry, bounded by
inner and outer radii $r_i$ and $r_o$. Its equation of motions
are given by the Navier-Stokes equations:
\EQA
\partial_t {\bf u}+{\bf u}{\cdot}\nabla {\bf
u}&=&-\frac{1}{\rho}\nabla p+\nu \Delta {\bf u},
\nonumber\\
\nabla\cdot {\bf u}&=&0.
\label{equans}
\ENA
where $\rho$ is the density, ${\bf u}$ is the velocity, $\nu$ the
molecular viscosity and  $p$ is the pressure.\
If we assume hydrostatic equilibrium in the vertical direction, we get:
\begin{equation}
\partial_z p \simeq 0,
\label{hydro}
\end{equation}
so that $p$ is a function of $r$ only. In that case, equation
(\ref{equans}) admits simple basic state under the shape of
stationary solutions, with axial and translation symmetry along the
disk rotation axis (the velocity only depends on $r$). The
incompressibility condition then implies:
\begin{equation}
\frac{1}{r}\partial_r (r u_r)=0,
\label{incomp}
\end{equation}
or
\begin{equation}
u_r=\frac{K}{r},
\label{circula}
\end{equation}
where $K$ is a constant, to be constrained later.
Plugging this into the radial and azimuthal component of
(\ref{equans}), we obtain two equations:
\EQA
-\frac{K^2}{r^3}-\frac{u_\theta^2}{r}&=&-\partial_r
\frac{\Pi}{\rho},\nonumber\\
\frac{K}{r^2}\partial_r(r
u_\theta)&=&\frac{\nu}{r^2}\partial_r\left(r^3\partial_r\frac{u_\theta
}{r}\right).
\label{equatheta}
\ENA
The second equation of (\ref{equatheta}) is homogeneous in $r$. It
only admits two power law solutions, with exponent $-1$ and
$1+K/\nu$, so that the general solution is:
\begin{equation}
u_\theta=A r^{1+\beta}+\frac{B}{r},
\label{solutheta}
\end{equation}
where $A$ and $B$ are constants and $\beta=K/\nu$. Plugging this
solution into the first equation of (\ref{equatheta}) then defines
the pressure.\

The basic state depends on three constants $A$, $B$ and $K$, which
must be specified through boundary conditions. In laboratory flows,
these conditions are usually well defined and allow for a simple
determination of the constants once the rotation velocities at the
inner and outer boundaries are known (Bahl, 1970):
\EQA
A&=&\frac{r_o^{-\beta}}{1-\eta^{\beta+2}}\left(\Omega_o-\eta^2\Omega_i
\right),\nonumber\\
B&=&\frac{r_i^2}{1-\eta^{\beta+2}}\left(\Omega_i-\Omega_o\eta^\beta\right),
\label{CLGeneral}
\ENA
where $\eta=r_i/r_o$ is the radius ratio, $\Omega_o$ and $\Omega_i$
are the angular velocity at outer and inner radii and
$\beta=R_r=K/\nu=u_r(r_i)r_i/\nu$ is the radial Reynolds number,
based on the radial velocity through the wall of the inner cylinder.
Note that it is positive for motions outwards from the axis of
rotation. \

Comparing (\ref{solkep}) with (\ref{CLGeneral}) and
(\ref{solutheta}), it is possible to see that the "laboratory" analog
of Keplerian flow is such that:
\EQA
\beta&=&-\frac{3}{2},\nonumber\\
\Omega_i&=&\Omega_\ast,\nonumber\\
\Omega_o&=&\sqrt{\frac{GM}{r_o^3}},\nonumber\\
\eta&=&r_i/r_o.
\label{limitdisk}
\ENA
This shows that this
basic state describes as well laboratory incompressible flows, with
rigid boundaries and without gravity, in which angular momentum
distribution may be imposed by boundary conditions, and astrophysical
flows, without rigid boundaries, in which angular momentum
distribution is imposed by gravity. In other words, building a
prototype of astrophysical disks (within the approximations described
above) requires a (laboratory) flow with equivalent angular momentum
distribution, and equivalent control parameters. We now derive these
control parameters, in order to apply the Reynolds similarity
principle.

\subsection{Control parameters}
The shape of the basic state allows for the determination of the
control parameters of the flow. These parameters are essential in the
comparison with the laboratory prototype since the Reynolds
similarity principle states that the astrophysical disk will behave
like the laboratory prototype with same control parameters. These
control parameters are \\
the global Reynolds number:
\begin{equation}
Re=\frac{\bar S (r_o-r_i)^2}{\nu},
\label{Reynolds}
\end{equation}
the rotation number:
\begin{equation}
R_\Omega=\frac{2\bar \Omega}{\bar S},
\label{Rotation}
\end{equation}
the curvature number
\begin{equation}
R_C=\frac{\bar r}{r_o-r_i},
\label{Curvatuire}
\end{equation}
the local radial Reynolds number:
\begin{equation}
R_r=\frac{u_r r}{\nu}.
\label{Radial}
\end{equation}
the aspect ratio:
\begin{equation}
\Gamma=\frac{\bar H}{\bar r}.
\label{aspect}
\end{equation}
Here, $\bar \Omega$, $\bar S$ and $\bar r$ are characteristic angular
velocity, shear and radius. Adopting the convention of Dubrulle et al
(2004b), we find:
\EQA
\bar \Omega
&=&\Omega_K(r_i)\left(\frac{r_i}{r_o}\right)^{1/2}+\frac{r_i}{r_o}\left( 
\Omega_\ast-\Omega_K(r_i)\right)\nonumber\\
\bar S&=&\frac{1}{2}\Omega_K(r_i)\left(\frac{r_i}{\bar
r}\right)^{3/2}-2\bar \Omega,\nonumber\\
\label{defibar}
\ENA
while $\bar r$ is fixed through the condition $\bar
\Omega=u_\theta(\bar r)/\bar r$. A simplification occurs in two
limiting cases, relevant to astrophysical disk: $r_i\ll r_o$ or
$r_i\to r_o,\quad \Omega_\ast \to \Omega_K(r_i)$. In both cases,  we
have:
\EQA
\bar r &=& r_i^{2/3} r_o^{1/3},\nonumber\\
\bar \Omega&=& \Omega_K(\bar r),\nonumber\\
\bar S&=&-\frac{3}{2}\Omega_K(\bar r).
\label{simpli}
\ENA
The control parameters then simplify into:
\EQA
Re&=& \frac{3}{2}\frac{\Omega_K(\bar r)(r_o-r_i)^2}{\nu},\nonumber\\
R_\Omega&=&-\frac{4}{3},\nonumber\\
R_C&=&\frac{r_i^{2/3} r_o^{1/3}}{r_o-r_i},\nonumber\\
R_r&=&-\frac{3}{2},\nonumber\\
\Gamma&=&\frac{H}{r_i^{2/3} r_o^{1/3}}.
\label{controlouf}
\ENA

If the disk is stratified or magnetized, other control parameters
appear, like the Prandtl number $Pr=\nu/\kappa$ and the magnetic
Prandtl number $Pm=\mu_0\nu/\eta$ where $\kappa$ and $\eta$ are the heat
diffusivity and the magnetic resistivity, and $\mu_0$ is the vacuum permeability.

\subsection{Molecular transport processes}

The computation of the control parameters requires an estimate of
molecular transport coefficients. These coefficients depend on the
ionization state of the gas. There are two sources of ionization.
Thermal ionization is efficient in the inner part of the disk. The
corresponding ionization fraction can be written as (Fromang et al,
2003):
\begin{equation}
x_e^{th}=6\times 10^{-2} \left(\frac{T}{1000K}\right)^{3/4}
\left(\frac{2.4\times 10^{15} cm^{-3}}{n}\right)^{1/2}\exp(-25188\ K/T),
\label{xeth}
\end{equation}
where $T$ and $n$ are the temperature and the number density of
neutral species (hydrogen mainly), respectively. For temperature lower than $10^3$
K (typically $r>1$ A.U.), thermal ionization is negligible. However,
X-Ray
illumination from the central star
(or the magnetospheric accretion flow)
may induce a weak ionization in
some part of the disk (typically away from the mid-plane) (Feigelsson
and Montmerle, 1999). A recent theoretical study has recently been
performed by Igea and Glassgold (1999). They found that at a given
radius from the source, the ionization fraction is a universal
function of the vertical column density $N_\perp$, independent of the
structural details of the disk. The role of cosmic rays in the disk
ionization is still a matter of debate (Sano et al. 2000) and will not be discussed here. For a typical young stellar object,
Igea and Glassgold's result can be approximated by:
\EQA
x_e^{X}&=&\frac{n_e}{n}=\frac{10^{17}\;{\rm cm^{-2}}}{N_\perp}T^{1/4}
n^{-1/2}e^{-0.002 (r_{AU}-1)},\quad N_\perp>10^{20} {\rm
cm}^{-2},\nonumber\\
x_e&=&\frac{n_e}{n}=10^{-3}T^{1/4} n^{-1/2}e^{-0.002 (r_{AU}-1)},\quad
N_\perp<10^{20} {\rm cm}^{-2},
\label{ionization}
\ENA
where $n_e$ is the ion number density, and $r_{AU}$
is the distance from the central star, in astronomical units.\

When the gas is neutral, the viscosity and heat diffusivity are given
by (Lang, 1980):
\begin{equation}
\nu_{neu}=\kappa=3\times 10^{19}\frac{T^{1/2}}{n}cm^2 s^{-1},
\label{visco}
\end{equation}
where $T$ is the disk temperature, and $n$ the number density. \

When the gas is weakly ionized ($x_e\ll 1$), the transport
coefficient must be multiplied by a factor (Lang, 1980):
\begin{equation}
\frac{\nu_{ion}}{\nu_{neu}}=4\times 10^{-12} \frac{T^2}{x_e}.
\label{correction}
\end{equation}
This correction is valid as long as $\nu_{ion}/\nu_{neu}<1$.
This sets a limit on
ionization fraction, below which the gas viscosity takes the neutral value:
\begin{equation}
x_e^{cr}=4\times 10^{-12}T^2.
\label{xecr}
\end{equation}
The Prandtl number in this case is equal to $10^{-11}$ (Lang, 1980).\

The resistivity of an ionized gas can be written as the sum of the
resistivity induced by electron-neutral collisions and electron-ion collisions:
\begin{equation}
\eta = \eta_{en} + \eta_{ei}
\end{equation}

where (Lang, 1980)

\begin{eqnarray}
\eta_{en} &=& 10^{-6}\frac{1-x_e}{x_e} T^{1/2} ohm-cm \\
\eta_{ei} &=& 4 \times 10^3 \frac{ln\ \Lambda}{T^{3/2}}\ ohm-cm
\end{eqnarray}
where $\Lambda=1.3 \times 10^4 \frac{T^{3/2}}{N_e^{1/2}}$ is the Coulomb
logarithm.


\subsection{Physical parameters}
Various physical parameters are required to estimate the control
parameter.

Parameters associated with the disk are $r_o$, $r_i$, $\Gamma$ and
$\nu$.
The disk inner
radius depends on whether the disk/star interaction is direct or
indirect. In the first case, $r_i=r_\ast$. In the
second case, $r_i$ may no exceed the corotation radius, at which the
disk velocity matches the star velocity. In the sequel, we shall
consider variations of $r_i$ in between these two limits.
The disk outer radius $r_o$ must be specified through the implicit
relation (\ref{defiro}). Direct observation for disk suggest that the
disk size is of the order of $r_D=1000$ A.U. for disk around T Tauri
and even smaller $r_D=50$ A.U.   for disk around FU ORI (Kenyon,
1999). Clearly, $r_D$ is thus the maximum size $r_o$ can achieve. For
practical reasons, we defer its discussion in the next section, after
computation of the temperature and density profile.\

Parameters relative to
the proto-star have been measured in some T Tauri and FU Ori stars.
Table \ref{tab:parameterstars} gives a sample of stars we shall
use in the following. It is particularly convenient and illustrative
to scale all quantities in the problem with respect to values defined
at the distance of $\bar r=0.33$ A.U., which is the characteristic
radius corresponding to a disk with $r_{i}=10^{11}$ cm and $r_o=1000$
A.U., the two extreme limits for $r_i$ and $r_o$. We also use, as a
reference, a rotation period of
the star of $8$ days (typical T Tauri star), leading to
$\Omega_\ast=9\times 10^{-6}$ s$^{-1}$.\

Temperature and number density in circumstellar disks are
not known due to the lack of
spatial resolution. However, their magnitude can
be deduced by short radii extrapolation of measurements made on the outer
disk. The inversion method of Guilloteau \& Dutrey
(1998), based on $\chi$-square fitting of CO interferometric maps, yields
the temperature and the density profiles at $r \gtrsim 50-100 $ A.U..
For instance, for the disk around DM Tau ($M \simeq 0.5 M_\odot$),
their method predicts
\[
\left\{
\begin{split}
T & \simeq 30 \left( \frac{r}{100 \; {\rm AU}} \right)^{-0.6} \qquad {\rm K}\\
n & \simeq 10^8 \left( \frac{r}{100 \; {\rm AU}} \right)^{-2.75} \qquad
{\rm cm}^{-3}\\
\Sigma & \simeq 1 \left( \frac{r}{100 \; {\rm AU}} \right)^{-1.5} \qquad
{\rm g\;cm}^{-2}, \\
\end{split}
\right.
\]
where $\Sigma$ is the surface density of the disk (including gas and dust).
Error bars on the measurements are rather large, and could amount to
a possible variation by a factor 5 to 10.
At $r=1$ A.U., one finds $n \simeq 3\times 10^{13}$ cm$^{-3}$ and
$\Sigma\simeq 1000$ gcm$^{-2}$, in agreement with values 
obtained by
modeling the deuterium enrichment in the Solar System (Drouart et
al. 1999; Hersant, Gautier \&  Hur\'e 2001).
At the reference radius of $0.33$ A.U., the density is $\bar n
\simeq 7 \times 10^{14}$ cm$^{-3}$, the surface density is
$\bar \Sigma\simeq 5275$ gcm$^{-2}$, the height is $\bar H=4.6\times
10^{12}$ cm and the temperature
is $\simeq 925$ K.
In a more recent analysis of the disk around BP Tau, Dutrey at al.
(2003) found densities and temperature
corresponding to a value of $\bar n \simeq 3 \times 10^{14}$
cm$^{-3}$, $\bar \Sigma\simeq 992$ g cm$^{-2}$, $\bar H=10^{11}$ cm
and $\bar T=289$ K. The difference between these figures and the
figures of DM Tau provide an illustration of the error bars
associated with our "typical values", since the disk
around BP Tau seems much smaller, and correspond to a more evolved
stage than the disk around DM Tau.\

\subsubsection{Ionization state}

It is interesting to study the ionization state of the disk with
temperature and density observed in DM Tau. The ionization
fraction is plotted as a function of radius in Fig.
(\ref{fig:ionize.eps}) for the thermal and X-ray contribution. One
sees that the thermal contribution dominates at $r<1$ A.U., while the
X-Ray contribution becomes important at larger radii. However,
comparing with the limiting ionization state eq. (\ref{xecr}), one
sees
that only the outer part of the disk $r>100$ A.U. is sufficiently
ionized to influence the molecular viscosity.

\begin{figure}[hhh]
\centering
\includegraphics[width=8.5cm]{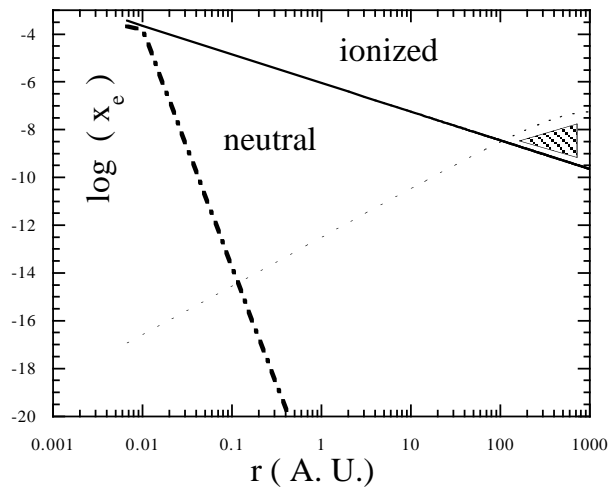}
\caption[]{Midplane
ionization fraction due
to thermal contribution (dot-dashed line) and X-ray contribution
(dot-dotted line). The plain line is the limiting ionization
fraction, below which the ionization does not influence molecular
transport. The shaded area is the region where ionization has to be
taken into account in the computation of the viscosity.}
\label{fig:ionize.eps}
\end{figure}

%

\subsection{Regimes}
The previous scaling allows for an estimate of the disk control
parameter, and, therefore, for an identification of the possible regimes.
A difficulty with respect to the laboratory experiment is that in
disks, the transport coefficients vary over across the disk due to
the radial stratification. To define the control parameter, one must
pick up a typical value. In this context, it is logical to consider
their value at $r=\bar r$, since both the typical shear and radii at
this location have been used.
Using the values given in the previous section, we then obtain:
\EQA
Re & = & 2\times 10^{25}\left(\frac{M}{M_\odot}\right)^{1/2}
\left(\frac{\bar n}{7 \times 10^{14} \; {\rm cm}^{-3}}\right)
\left(\frac{\bar T}{930 \; {\rm K}}\right)^{-1/2}\nonumber\\
     &&\qquad \times\left(\frac{r_{i}}{10^{11} \;{\rm
cm}}\right)^{-4/3}\left(\frac{r_o}{10^3 \;{\rm
A.U.}}\right)^{-2/3}\,\nonumber\\
R_\Omega&=&-\frac{4}{3},\nonumber\\
R_r&=&-\frac{3}{2},\nonumber\\
Pr&=&1,\nonumber\\
Pm&=&2\times 10^{-8},\nonumber\\
\Gamma&=&0.94.
\label{controlouf}
\ENA
The value of the rotation number indicates that the flow is
anti-cyclonic and belongs to the "globally sub-critical" class
defined in Dubrulle et al (2004b). The radial Reynolds number is
negative, indicating an inward
radial circulation. Its value is close to unity. So its influence on transport
properties can be neglected as a first approximation, see Dubrulle et
al (2004b). The curvature number depends on the interpretation of the 
viscous time scale (see Section 2.9.2). In the case where the 
interpretation is done with the molecular viscosity, one finds $ 
R_C=0.0004$, that is disks are in the wide gap limit. In the other 
limiting case where the viscous time scale is computed using the 
turbulent viscosity, one finds $R_C=1-\Gamma$, that is disks are in 
the small gap limit.
The relevant
Reynolds parameter to be considered in studying transport properties
will be $ReR_C^2=\bar S \bar r^2/\nu$, see Dubrulle et al (2004b).
Finally, the aspect ratio is less than one. From the review of
Dubrulle et al, we infer that an additional correction $\Gamma^{2}$
must be
included in the definition of the relevant Reynolds parameter, which
becomes:
\EQA
Re_{phys}&=&Re (R_C \Gamma)^2=\frac{\bar S {\bar H}^2}{\nu},\nonumber\\
&=& 3\times 10^{13}\left(\frac{M_\ast}{M_\odot}\right)^{-1/2}
\left(\frac{\bar n}{7 \times 10^{14} \; {\rm cm}^{-3}}\right)
\left(\frac{\bar T}{930 \; {\rm K}}\right)^{1/2}\nonumber\\
     &&\qquad \times\left(\frac{r_{\rm in}}{10^{11} \;{\rm
cm}}\right)^{-1}\left(\frac{r_o}{10^3 \;{\rm
A.U.}}\right)^{-1/2}\,
\label{releventRe}
\ENA
This is the expression one would naturally derive by
considering the "smallest" length scale in the problem, see e.g.
Longaretti (2003).

\section{Predictions about the structure of circumstellar disks}
\subsection{Stability: the laminar/turbulent transition}
The stability properties of circumstellar disks can be found by
comparing the physical Reynolds number $Re_{phys}$ with critical
Reynolds numbers
derived in laboratory experiments, in the anti-cyclonic non-linear
regime. These measurements are summarized in Dubrulle et al (2004b).
Disregarding any body forces, one finds a critical Reynolds number of
the order of 2300, well below the disk value. Taking into account the
possible stable vertical stratification observed e.g. in DM Tau (Dartois et
al, 2003) \footnote{This may be due to disk illumination by the
central star (D'Alessio et al, 1998)}, one obtains a slightly larger
value of the order of $4000$ (Dubrulle et al, 2004a). The presence of
a vertical magnetic field may increase the critical Reynolds number,
due to the low magnetic Prandtl number prevailing in disks. Using the
scaling of Willis and Barenghi (2002), one finds a critical Reynolds
number $Re_c\sim 100/Pm=10^{10}$. This is still well below the
observed Reynolds number. These number concur to conclude that the
disk is turbulent.\

However, due to the huge variation of the transport coefficients
across the disk, one may wonder how strong this conclusion is. A way
to answer this question is to see how {\sl locally in the disk}, the
stability criterion are satisfied using "local" non-dimensionalized
parameter, built by replacing $\bar r$ by $r$ the distance to the
central object. The result of this procedure is plotted on Fig
\ref{fig:critique.eps}. One sees that at any radius, such effective
local Reynolds number is well above any critical Reynolds number due
to body forces.
This strengthens our conclusion.

\begin{figure}[hhh]
\centering
\includegraphics[width=8.5cm]{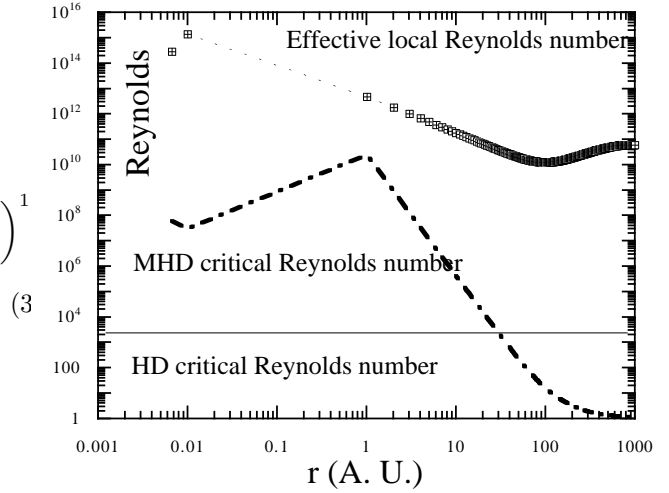}
\caption[]{Physical local Reynolds number in circumstellar disks as a
function of radius (dotted-line with symbols). The dashed-dotted line
and the full line are the
critical Reynolds number deduced from laboratory experiments, see
Dubrulle et al (2004b).}
\label{fig:critique.eps}
\end{figure}

\subsection{Mean energy dissipation and accretion rate}
\label{subsec:energydissipation}
\subsubsection{Definition}
The quantitative comparison between experimental
measurements and energy dissipated in
circumstellar disks first requires a relation linking
      the torque  and the disk luminosity ${\cal L}$.
The total power dissipated in a Taylor-Couette experiment with same
control parameter as in a keplerian disk is
\begin{equation}
\epsilon=\bar \nu^2\bar \Sigma \frac{\bar S G}{4}=\frac{G}{4
Re_{phys}^2}\bar \Sigma \bar S^3 \bar H^4 ,
\label{eq:defepsilon}
\end{equation}
where $\bar \Sigma$ is the disk surface density, $G$ the non-dimensional
torque and $\bar \nu$ the viscosity.
In a stationary disk, this
power is dissipated under the form of heat, and thus coincides with
the disk luminosity, that is
\begin{equation}
{\cal L}=\epsilon.
\label{eq:lisepsilon}
\end{equation}
In practice, observed luminosities are often expressed as a function
of an "effective mass accretion rate", namely (Hartmann et al. 1998)
\begin{equation}
\dot M= \frac{0.8 r_\ast {\cal L}}{{\cal G} M}.
\label{eq:defmdot}
\end{equation}
    From a theoretical point of view, the detailed computation of the
accretion luminosity is
not straightforward since it depends on the boundary condition at the
interaction radius. For comparison with experimental data, we
therefore consider the quantity:
\begin{equation}
\frac{G}{Re_{phys}^2}=\frac{\dot M}{\dot M_0},
\label{considerons}
\end{equation}
where $\dot M_0$ is an effective accretion rate given by
\begin{equation}
\dot M_{0}=\bar \Sigma r_\ast \bar r\bar \Omega\left(\frac{\bar
H}{\bar r}\right)^4.
\label{utileMast}
\end{equation}
The quantity $G/Re_{phys}^2$ (the non-dimensional energy dissipation),
includes all the boundary condition dependence and only depends on
the Reynolds number.

\subsubsection{Prediction using laboratory experiments}
The result of Dubrulle et al (2004b) lead to an analytical
prediction for the function $\dot M/\dot
M_0=f(R_\Omega)G_i/Re_{phys}^2$, where $f(R_\Omega)$ is a function
parameterizing the influence of rotation, and $G_i$ is the torque in
situation when only the inner cylinder is rotating. From the
experiments, we infer $f(R_\Omega)\sim 0.1$
if the flow is turbulent, and $1$ is the flow is laminar.  Taking the
wide gap limit $r_i\ll r_o$, $R_C^2 Re=Re_{phys}$ in the formulae of
Dubrulle et al (2004b), we obtain three possible regimes:
\label{note:re}
\begin{itemize}
\item in the laminar regime, for $Re_{phys} \le 2300$
\begin{equation}
\frac{\dot M}{\dot M_{0}}
= \frac{2\pi}{Re_{phys}}.
\label{nuequiv}
\end{equation}
\item for smooth boundary conditions and $ Re_{phys}>2300$
\begin{equation}
\frac{\dot M}{\dot M_{0}}\vert_{smooth}
=0.06 \left[ \ln( 3\times 10^{-4}  Re_{phys}^2
      )\right]^{-3/2}.
\label{regime3taylor}
\end{equation}
\item for rough boundary conditions and $ Re_{phys}>2300$
\begin{equation}
\frac{\dot M}{\dot M_0}\vert_{rough}=1.9\times 10^{-2}.
\label{inertial}
\end{equation}
\end{itemize}
In the other limit $r_i\to r_o$, the turbulent value take similar 
expression, but must be multiplied by a factor $1/3(1-\Gamma)^{3/2}$.
In astrophysical disks, the boundary conditions are not known a
priori. Moreover, given the huge physical difference between the
inner part and the outer part, it is unlikely that the boundary
condition at the inner and outer part coincide, so that we are
probably more in a state of "mixed" boundary conditions studied
experimentally by Van den Berg et al (2003). In that case, the energy
dissipation is found to vary in between the two limits set by
respectively the "pure" smooth type (\ref{regime3taylor}) and the
pure "rough" type (\ref{inertial}), see Dubrulle et al (2004b). We
shall therefore adopt these formulae as a lower and upper limit of
the energy dissipation in disks.

\begin{figure}[hhh]
\centering
\includegraphics[width=8.8cm]{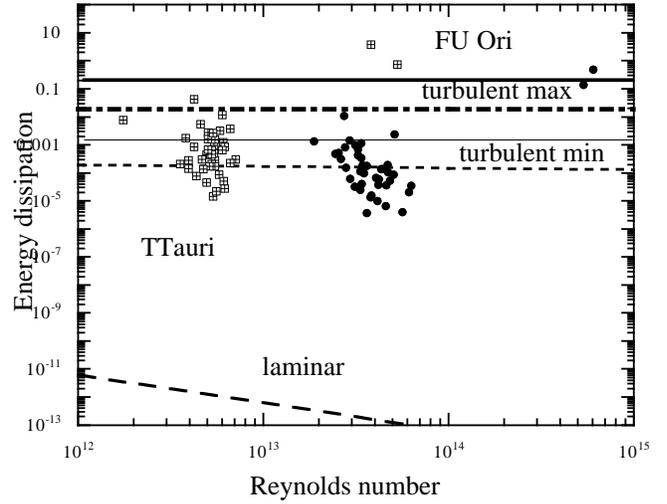}
\caption[]{Comparison between the non-dimensional energy dissipation
predicted from laboratory measurements in the wide gap limit ({\it 
dotted lines}) or in the small gap limit ({\it plain lines}) and 
observed in
circumstellar disks ({\it symbols}), as a function of
the Reynolds number $Re_{phys}$. For an easier comparison, the mean
energy dissipation has been translated into the non-dimensional
accretion rate $\dot{M}/\dot{M}_{0}$
(computed using  Eq.(\ref{regime3taylor}) and observationally
determined parameters reported in Tab. \ref{tab:parameterstars}). The 
symbols $\boxplus$
    report the value using $r_{i}=r_{\rm coro}$, which provides
an upper bound of the energy dissipation. The 
circles
report the value using $r_{i}=r_\ast$, which provides the lower
bound of the energy dissipation. All the quantities have been
computed using temperature and density estimated for the DM Tau
system using the results of Guilloteau and Dutrey (1998), so there is
no adjustable parameter in this plot.}
   \label{fig:compaastrolabo}
\end{figure}

\subsubsection{Test against observational data}
\label{sec:Model predictions and implications}

We find the following scaling for $\dot M_{0}$:
\EQA
&&\dot M_{0}  \simeq  3 \times 10^{-5}\left (\frac{\bar\Sigma}{5300
\; {\rm gcm}^{-2}}\right)
\left(\frac{M}{M_\odot}\right)^{-3/2}
\left(\frac{r_\ast}{10^{11} \;{\rm
cm}}\right)\nonumber\\
     &&\qquad \times\left(\frac{r_{i}}{10^{11} \;{\rm
cm}}\right)^{-0.8}\left(\frac{r_o}{10^3 \;{\rm
A.U.}}\right)^{-0.4}\,\left(\frac{\bar T}{930 \; {\rm K}}\right)^{2}
\; {\rm M}_\odot{\rm /yr}.
\label{utileMast2}
\ENA
We are aware that these values are probably uncertain by a factor of
$10$ or even more. In the next decade, the results expected with ALMA
    will largely shorten the error bars.

\begin{table*}
\centering
\hspace*{-0.5cm}\begin{tabular}{|c|ccccc|cccc|} \hline
         & \multicolumn{5}{|c}{input parameters} &
\multicolumn{4}{|c|}{output parameters}\\ \hline \hline
          & mass    & radius &coro. radius & disk outer edge & accr. 
rate  &  \multicolumn{2}{|c}{eff. rates}
&\multicolumn{2}{c|}{Reynolds num.} \\
star & $M/M_\odot$ & $r_{*}/R_\odot$ & $r_{\rm coro}/R_\odot$ & 
$r_o/1000$ A.U. & $\log
\dot{M} $  & $\log \dot M_{0}(r_\ast)
$& $\log \dot M_{0}(r_{corot}) $  &$\log Re_{\ast}$ &$\log Re_{\rm coro}$ \\
\hline
AA Tau &0.53 &1.74 &13.81 &1 &-8.48 &-4.39 &-3.67 &13.53 &12.55\\
BP Tau &1.32 &1.99 &17.79 &1 &-7.54 &-2.87 &-2.10 &13.27 &12.24\\
CY Tau &0.42 &1.63 &10.05 &1 &-8.12 &-4.17 &-3.54 &13.61 &12.74\\
DE Tau &0.25 &2.45 &10.05 &1 &-7.58 &-4.00 &-3.52 &13.54 &12.85\\
DF Tau &0.27 &3.37 &11.29 &1 &-6.91 &-3.31 &-2.90 &13.39 &12.78\\
DK Tau &0.43 &2.49 &10.05 &1 &-7.42 &-3.49 &-3.01 &13.42 &12.73\\
DN Tau &0.38 &2.09 &10.03 &1 &-8.46 &-4.60 &-4.06 &13.52 &12.76\\
DO Tau &0.37 &2.25 &10.05 &1 &-6.84 &-3.00 &-2.48 &13.50 &12.77\\
DQ Tau &0.44 &1.79 &10.05 &1 &-9.40 &-5.43 &-4.83 &13.56 &12.73\\
DS Tau &0.87 &1.36 &10.05 &1 &-7.89 &-3.45 &-2.76 &13.53 &12.58\\
GG Tau &0.44 &2.31 &10.05 &1 &-7.76 &-3.81 &-3.30 &13.45 &12.73\\
GI Tau &0.71 &1.48 &10.05 &1 &-8.02 &-3.72 &-3.06 &13.53 &12.63\\
GK Tau &0.46 &2.15 &10.05 &1 &-8.19 &-4.21 &-3.67 &13.47 &12.72\\
GM Aur &0.52 &1.78 &10.05 &1 &-8.02 &-3.95 &-3.34 &13.52 &12.69\\
HN Tau &0.81 &0.76 &10.05 &1 &-8.89 &-4.45 &-3.55 &13.80 &12.60\\
IP Tau &0.52 &1.44 &10.05 &1 &-9.10 &-5.00 &-4.33 &13.62 &12.69\\
UY Tau &0.42 &2.60 &10.05 &1 &-7.18 &-3.27 &-2.80 &13.41 &12.74\\
CI Tau &0.5 &1.87 &10.05 &1 &-7.19 &-3.14  &-2.56 &13.51 &12.70\\
CX Tau &0.33 &1.63 &10.05 &1 &-8.97 &-5.18 &-4.55 &13.66 &12.79\\
CZ Tau &0.41 &1.19 &10.05 &1 &-9.35 &-5.39 &-4.65 &13.75 &12.74\\
DM Tau &0.43 &1.39 &10.05 &1 &-7.95 &-3.97 &-3.29 &13.67 &12.73\\
DD Tau &0.42 &1.44 &10.05 &1 &-8.39 &-4.43 &-3.76 &13.66&12.74\\
DH Tau &0.38 &1.67 &11.33 &1 &-8.30 &-4.42 &-3.76 &13.62 &12.71\\
DI Tau &0.43 &1.71 &12.56 &1 &-8.75 &-4.79 &-4.01 &13.58 &12.64\\
DP Tau &0.46 &1.44 &10.05 &1 &-7.88 &-3.86 &-3.19 &13.64 &12.72\\
FM Tau &0.58 &1.17 &10.05 &1 &-8.45 &-4.26 &-3.52 &13.68 &12.67\\
FO Tau &0.33 &1.59 &10.05 &1 &-7.50 &-3.71 &-3.07 &13.67 &12.79\\
FQ Tau &0.35 &1.42 &10.05 &1 &-6.45 &-2.61 &-1.93 &13.71 &12.78\\
FS Tau &0.46 &1.25 &10.05 &1 &-8.09 &-4.06 &-3.34 &13.70 &12.72\\
FV Tau &0.71 &1.87 &10.05 &1 &-6.23 &-1.95 &-1.37 &13.44 &12.63\\
FX Tau &0.34 &1.94 &10.05 &1 &-8.65 &-4.86 &-4.28 &13.58 &12.79\\
FY Tau &0.50 &1.87 &10.05 &1 &-7.41 &-3.36 &-2.78 &13.51 &12.70\\
GH Tau &0.29 &1.90 &10.05 &1 &-7.92 &-4.23 &-3.65 &13.62 &12.82\\
GO Tau &0.50&1.40 &10.05 &1 &-7.93 &-3.86 &-3.17 &13.64 &12.70\\
Haro 6-37 &0.60 &1.90 &10.05 &1 &-7.00 &-2.83 &-2.26 &13.46 &12.66\\
HO Tau &0.56 &0.94 &10.05 &1 &-8.86 &-4.68 &-3.86 &13.79 &12.68\\
IQ Tau &0.35 &2.01 &10.05 &1 &-7.55 &-3.74 &-3.18 &13.56 &12.78\\
LkCa 15 &0.81 &1.53 &10.05 &1 &-8.87 &-4.49 &-3.84 &13.49 &12.60\\
Lk Ha 332/G1 &0.29 &2.36 &10.05 &1 &-6.60 &-2.93 &-2.43 &13.53 &12.82\\
V955 Tau &0.44 &2.34 &10.05 &1 &-7.02 &-3.07 &-2.57 &13.44 &12.73\\ \hline
V1057 Cygni &0.50 &1.60 &13.53 &0.005 &-4.00 &-0.86 &-0.12 &14.73 &13.72\\
FU Ori &0.70 &1.20 &15.83 &0.005 &-3.70 &	-0.31 &-0.58 &14.78 &13.58\\
\hline
\end{tabular}
\caption{Observational parameters for T Tauri stars (from  Bouvier
1990, Hartmann et al. 1998)
     and for FU Ori stars (from Popham et al. 1996)
     considered in this study ({\it left}) and disk physical parameter
({\it right}). The computation of the corotation radius requires the
knowledge of the star rotation velocity. In case this last quantity
is not available, the corotation radius has been set to $10.05$, the
solar value.  Lower and upper bound on $\dot M_0$ and $Re$ have been
computed using either the star radius or the corotation radius for
$r_{i}$. Accretion rates are in M$_\odot$yr$^{-1}$.}
\label{tab:parameterstars}
\end{table*}

At $Re=3\times 10^{13}$, our model (Eqs. (\ref{regime3taylor}) and
(\ref{inertial})  predicts that $0.0002<\dot M/\dot M_0<0.019$,
resulting in
$6\times 10^{-9}<\dot M<6\times 10^{-7} M_\odot$yr$^{-1}$ for T Tauri
stars. This is in good agreement with the observed values ranging
from ${\dot M}=10^{-10}$
to $10^{-6} M_\odot$yr$^{-1}$ (Hartmann et al. 1998). Disks around FU
Ori are characterized by a smaller disk radius, leading to higher
values of $\dot M$
according to (\ref{utileMast2}), by a factor $5$. This is not quite
enough to reach values of up to $10^{-5} M_\odot$ yr$^{-1}$,
associated to
disks around
FU Ori stars (Kenyon 1994). Such values could be obtained if the
typical disk density is higher in disk around FU Ori, resulting in
more massive disks. This is plausible, since FU Ori are younger than
T-Tauri stars.\

A graphical representation of this discussion can be obtained by
plotting the computed $\dot M/\dot M_0$ as a function of $Re_{phys}$
using the values listed in Table \ref{tab:parameterstars} as input
parameters. To remove the problem with our ignorance of the actual
value of $r_{i}$, we have used the relation $r_\ast\le r_{i}
\le r_{\rm coro}$ and computed the corresponding $\dot M/\dot M_0$
and $Re_{phys}$. The actual dissipation somehow lies in between the
two corresponding estimates. These estimates are plotted in
Figure \ref{fig:compaastrolabo}. For comparison, we have added the
theoretical predictions (eq. (\ref{regime3taylor}) and
(\ref{inertial}) giving the minimum and the maximum expected values
in the turbulent case, as well as the laminar value. This last value
is very much lower than the turbulent values, and is never even
nearly approached by any stars we considered. This may be seen as a
proof of the turbulent character of all these disks.\

We also see that energy dissipation for disk around T Tauri has a
tendency to cluster in between the minimal and maximal values allowed
by the theoretical predictions. The relative position of the cluster
of point is slightly better in the case where $r_{i}$ is computed
with $r_{corot}$, which may be an indication that $r_{i}$ is actually
closer to the corotation radius than to the star radius. However,
given the error bars stressed above, this is probably not enough to
conclude that disk around T Tauri stars are connected through a
magneto-sphere, rather than through a boundary layer. In the case of
FU Ori, the points are clearly above the maximum allowed by our
choice of parameters. The discrepancy is slightly lower for the case
when $r_{i}=r_\ast$, a choice which should probably be favored by the
possible signatures of boundary layer in these objects (Kenyon,
1994). In that case, an increase of the surface density by a factor
10 with respect to our values will be enough to solve the discrepancy
with theory.\

Our estimate neglects the influence of the magnetic field. Laboratory
experiments using liquid metals have proved that this can potentially
change the intensity of the transport with respect to the pure
hydro-dynamical case. However, no experiment has been performed so
far, to study the magnetic influence in regimes relevant to
astrophysical disks.

\subsection{Turbulent viscosity}
The turbulent viscosity in circumstellar disks can be predicted by
comparison with laboratory measurements, see Dubrulle et al (2004b).
We find:
\begin{equation}
\nu_t=\frac{1}{2\pi}\frac{\tau_{lam}}{\bar
\tau}\frac{G}{Re_{phys}^2} \vert \bar S \vert \bar H^2,
\label{resultbefore}
\end{equation}
where $\tau=\Sigma S$ is proportional to the angular momentum, and
the index $lam$ means laminar value. The ratio $\tau_{lam}/\tau$ is a
function describing the radial variation of the turbulent transport.
In an incompressible laboratory flow with constant density, this
function is just the ratio of the laminar shear profile to the
turbulent shear profile: the turbulence regulates itself through a
modification of the velocity profile. In keplerian disks, the shear
profile is fixed (through the gravitational force), and the
regulation operates through the density. This function has been
measured in a number of laboratory experiments. At large Reynolds
number, it seems to approach a constant value of $4$ predicted by Busse
(1970) using argument of maximal momentum transport.
With  $\vert \bar S \vert =1.5 \Omega$ and $G/Re_{phys}^2=\dot M/\dot M_0$, this
defines a resulting typical turbulent viscosity as:
\begin{equation}
\bar \nu_t=\frac{3}{\pi}\frac{\dot M}{\dot M_0}\bar \Omega \bar H^2.
\label{resultbefore2}
\end{equation}
This turbulent viscosity takes the shape of an $\alpha$ viscosity,
proposed by Shakura and Sunyaev (1973). The corresponding $\alpha$
coefficient is here a function of the Reynolds number of the
circumstellar disk (through eqs (\ref{regime3taylor}) and
(\ref{inertial})). At $Re_{phys}=3\times 10^{13}$, its value is
typically $2\times 10^{-4}<\alpha<2\times 10^{-2}$. This range is in
good agreement with the range of values inferred from the D/H ratio
in the Solar System (Drouart et al, 1999; Hersant et al, 2001). More
generally, this range is compatible with disk lifetime and values
usually adopted in theories. Using this expression for the turbulent
viscosity, the viscous timescale writes :
\begin{equation}
t_\nu=\frac{\pi}{3} \frac{\dot M_0}{\dot M} \left(
\frac{r}{H}\right)^2 \frac{1}{\bar \Omega}
\end{equation}

\subsection{Energy fluctuations}
\label{subsec:fluctuations}

Up to now, we have considered only the mean energy dissipation and its
luminous counterpart, but we can also draw interesting information from the
     {\it luminosity fluctuations} which reflect the dynamics of the underlying
     turbulent flow. In laboratory experiments with smooth boundary
conditions, turbulent fluctuations are
     observed to follow an universal (i.e. Reynolds
number-independent), log-normal
distribution (Lathrop, Fineberg \& Swinney 1992) with variance $\Delta=0.042$.
The universal distribution occurs
for variable normalized by their mean. Energy dissipation is
proportional to the wall shear stress squared. Since the functional
shape of the log-normal distribution is unchanged by squaring,
distribution of energy dissipation should also be log-normal.
To check this prediction, we have computed the distribution of the
luminosity fluctuations observed from the disk around BP Tau and from
the disk around V1057 Cygni. The results are shown in Fig.
\ref{fig:BPTaufluc.ps} and \ref{fig:Vcygfluc.ps}. One sees that the
fluctuations in the disk around V1057 Cyg are very well fitted by a
log-normal distribution, with a variance similar to that of
laboratory experiments. In the case of BP Tau, however, the comparison
is not as good. This difference between the two systems may be traced
to different boundary conditions. If we accept that disk around
T Tauri stars follow the magnetospheric accretion scenario, while the disk
around FU Ori is connected to the star through a boundary layer, it may not be
surprising that only the disk around FU Ori follow the laboratory,
smooth boundary condition distribution. Since we do not have any
measurements for rough boundary conditions, we cannot say whether the
discrepancy comes from the different boundary conditions, or from the
presence of other physical effects, like accretion shock, or magnetic
field.

\begin{figure}[hhh]
\centering
\includegraphics[angle=270,width=9cm]{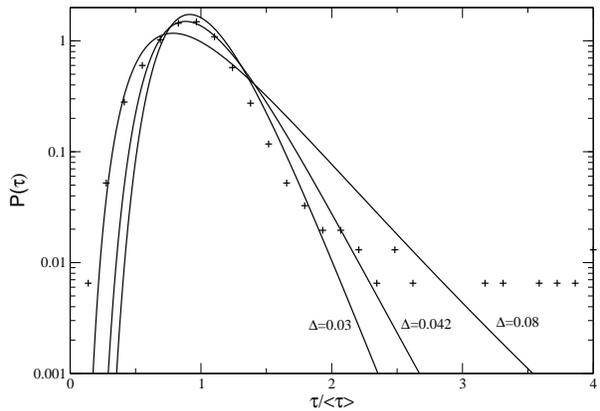}
\caption[]{Distribution of
luminosity fluctuations observed disk around BP Tau ({\it
symbols}) compared with a log-normal distribution of various variance
$\Delta$ ({\it plain line}). The value of $\Delta$ is indicated aside
each line.}
\label{fig:BPTaufluc.ps}
\end{figure}

\begin{figure}[hhh]
\centering
\includegraphics[angle=270,width=9cm]{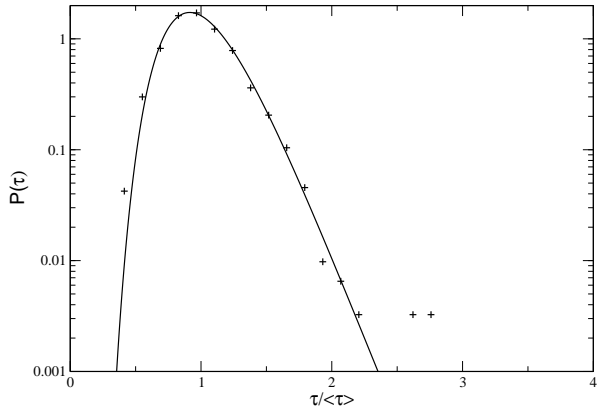}
\caption[]{Distribution of
luminosity fluctuations observed disk around V1057 Cyg ({\it
symbols}) compared with a log-normal distribution of variance
$\Delta=0.03$ ({\it plain line}).}
\label{fig:Vcygfluc.ps}
\end{figure}

\subsection{Velocity fluctuations}

Laboratory measurements provide interesting clues about the intensity
of velocity fluctuations. Since such fluctuations may be potentially
observable in disks using non-thermal line widening, they may be used
as additional constraints or observational test of the analogy
between laboratory flows and circumstellar disks. From results of
Dubrulle et al (2004b), it appears that azimuthal velocity
fluctuations should be proportional to the mean azimuthal velocity,
with a proportionality factor depending weakly on the Reynolds
number, like $0.03 (Re/Re_c)^{-0.125}$. With $Re_{phys}=3\times
10^{13}$ and $Re_c=10^8$ (see Section 3.1), the factor is of the
order of $0.01$.
Using Eq. (\ref{solkep}), we can compute the azimuthal velocity
dispersion for a typical circumstellar disk around a T Tauri.
The azimuthal velocity dispersion decreases from about 0.6 km/s in the
inner part, to 0.03 km/s in the outer part, at $100$ A.U. The total velocity
dispersion depends on the anisotropy of the turbulence. In the
laboratory experiment, the radial relative velocity dispersion is
observed to be about twice the azimuthal velocity dispersion. There
was no measure of the vertical velocity dispersion, but it can be
expected to be much smaller than the horizontal dispersion due to the
rotation-induced anisotropy (Dubrulle \& Valdettaro 1992).\

Velocity dispersion in disks have been measured by Guilloteau \& Dutrey (1998)
at $r>100$ A.U. They obtain a value of
the order of $0.1$ km/s,
which would correspond to a value of about $0.05$ km/s for the
azimuthal component.  This is close to the values found from
comparison with laboratory flows.

\section{Summary}

In this paper, we have derived and studied the analogy between
circumstellar disks and the Taylor-Couette flow. This analogy results
in a number of parameter-free predictions about stability of the
disks, and their turbulent transport properties, provided an estimate
of the disk structure is available. We have proposed 
to get this
estimate from interferometric observations of circumstellar disks, 
and checked that the energy dissipation, the turbulent
transport, and the fluctuations in circumstellar disks all follow
behavior compatible with the prediction from the analogy. This check
can first be used as a clear proof of the turbulent character
of circumstellar disks. A second interesting application would be to
build from this analogy a parameter free model of circumstellar
disks. In this respect, the proportionality between the turbulent
viscosity and the so called "accretion rate" (a quantity easily
accessible to observation) is very interesting because it opens the
possibility to {\sl infer} the disk structure from the observation of
its luminosity. For this, a model has to be built linking the
turbulent transport and the disk structure. This is the subject of
ongoing work.\

We note
finally that our model could also possibly apply to other type of
disks (e.g. around black holes, or in close binaries) provided
minor adaptations.

\begin{acknowledgements}
We thank A. Dutrey for useful discussions and comments on the
manuscript. We are indebted to the anonymous referee whose helpful remarks led us to clarify the paper and our thoughts. This work has received support from the Programme
national de Plan\'etologie. F.H. acknowledges support from an ESA
research
fellowship.
\end{acknowledgements}

\end{document}